\documentclass[fleqn,usenatbib]{mnras}

\usepackage{newtxtext,newtxmath}

\usepackage[T1]{fontenc}

\DeclareRobustCommand{\VAN}[3]{#2}
\let\VANthebibliography\thebibliography
\def\thebibliography{\DeclareRobustCommand{\VAN}[3]{##3}\VANthebibliography}

\usepackage{graphicx}	
\usepackage{amsmath}	

\title[Triggering thermonuclear micronovae]{Triggering micronovae through magnetically confined accretion flows in accreting white dwarfs}

\author[S. Scaringi et al.]{S. Scaringi,$^{1}$\thanks{E-mail:simone.scaringi@durham.ac.uk}
P.J. Groot,$^{2,3,4}$
C. Knigge,$^{5}$
J.-P. Lasota,$^{6,7}$
D. de Martino,$^{8}$
Y. Cavecchi,$^{9}$
\newauthor
D.A.H. Buckley,$^{3,4,10}$
and M.E. Camisassa$^{11}$
\\
$^{1}$Centre for Extragalactic Astronomy, Department of Physics, Durham University, South Road, Durham, DH1 3LE\\
$^{2}$Department of Astrophysics/IMAPP, Radboud University, P.O. 9010, 6500 GL, Nijmegen, The Netherlands\\
$^{3}$South African Astronomical Observatory, PO Box 9, Observatory, 7935, Cape Town, South Africa\\
$^{4}$Department of Astronomy, University of Cape Town, Private Bag X3, Rondebosch, 7701, South Africa\\
$^{5}$School of Physics and Astronomy, University of Southampton, Highfield, Southampton SO17 1BJ, UK\\
$^{6}$Nicolaus Copernicus Astronomical Center, Polish Academy of Sciences, ul. Bartycka 18, 00-716 Warsaw, Poland\\
$^{7}$Institut d’Astrophysique de Paris, CNRS et Sorbonne Universites, UMR 7095, 98bis Boulevard Arago, 75014 Paris, France\\
$^{8}$INAF-Osservatorio Astronomico di Capodimonte, Salita Moiariello 16, I-80131 Naples, Italy\\
$^{9}$Instituto de Astronomia, Universidad Nacional Autonoma de Mexico, Ciudad de Mexico, CDMX 04510, Mexico\\
$^{10}$Department of Physics, University of the Free State, PO Box 339, Bloemfontein, 9300, South Africa\\
$^{11}$Department of Applied Mathematics, University of Colorado, Boulder, CO 80309-0526, USA\\
}

\date{Accepted 2022 April 12. Received 2022 April 8; in original form 2022 March 7}

\pubyear{2022}

\begin{document}
\label{firstpage}
\pagerange{\pageref{firstpage}--\pageref{lastpage}}
\maketitle

\begin{abstract}
Rapid bursts at optical wavelengths have been reported for several accreting white dwarfs, where the optical luminosity can increase by up to a factor 30 in less than an hour fading on timescales of several hours, and where the energy release can reach $\approx10^{39}$ erg (``micronovae''). Several systems have also shown these bursts to be semi-recurrent on timescales of days to months and the temporal profiles of these bursts strongly resemble those observed in Type-I X-ray bursts in accreting neutron stars. It has been suggested that the observed micronovae may be the result of localised thermonuclear runaways on the surface layers of accreting white dwarfs. Here we propose a model where magnetic confinement of the accretion stream on to accreting magnetic white dwarfs may trigger localised thermonuclear runaways. The proposed model to trigger micronovae appears to favour magnetic systems with both high white dwarf masses and high mass-transfer rates. 

\end{abstract}



\section{Introduction}

Classical Novae (CN) are the result of thermonuclear runaways (TNRs) on the surface layers of accreting white dwarfs (AWDs). After the accumulation of hydrogen from a companion mass donor, ignition conditions are reached near the white dwarf (WD) surface, initiating a runaway thermonuclear explosion \citep[see e.g.][]{GS87}. Nova explosions result in an increase of up to 10 magnitudes or more at optical wavelengths, have rise times of days and remain bright for weeks to months \citep[see e.g.][]{warner03}. A subclass of novae, the recurrent novae (RN), are observed to show repeated outbursts on time scales of years to centuries. The shortest known recurrence time is one year in the system M31N 2008-12a, located in the Andromeda Nebula \citep[][]{darnley16}. The physics of novae is well understood, and outburst amplitude and recurrence times are directly related to the mass of the underlying WD and the mass-accretion rate from the donor star \citep[e.g.][]{starrfield72,SB07}. Crucially, a classical nova outburst is always a global phenomenon where the accreted hydrogen layer over the whole surface of the WD is burnt after a local ignition. An equivalent to novae in systems harbouring a neutron star accretor are the Type I X-ray bursts \citep[e.g.][]{lewin93}. Here the flame also ignites at one location and eventually covers the whole surface \citep[e.g.][]{SB06,GK21}. When a strong enough magnetic field is present, the flow is channeled on to a smaller fractional area on to the surface and there are indications that this may favour ignition at the base of the accretion column \citep[][]{goodwin21}.

Until recently, localised thermonuclear runaways (LTNRs) on AWDs have not been identified. \citet{M80} and \citet{shara82} proposed a mechanism that may allow LTNRs to occur on the surface of AWDs. This mechanism invokes transverse temperature gradients and inhomogeneities in the accreted layers that thermalise on timescales that are much longer than the thermonuclear runaway timescale. In this scenario the freshly accreted material will ignite and be consumed by the propagating flame. This model was initially developed to explain what had already been recognised as accretion-induced dwarf-nova (DN) outbursts \citep{Smak71,Warner74}. It is nonetheless interesting to note that both the rise times and the recurrence times of a LTNR expected from this model can be matched to those observed in DN outbursts of about $1$ day and several weeks respectively. \citet{OS93} later revised this model through numerical calculations to include the effects of mass-accretion and importantly the effects of meridional vs. radial energy transport. Extending the analytical model of \citet{shara82} to include these effects, \citet{OS93} demonstrate that conditions for triggering LTNRs are possible through temperature inhomogeneities, and are more likely to occur in systems accreting at high rates and on higher mass WDs. This would give rise to a non-spherical TNR, possibly explaining the observed asymmetries in some nova shells \citep[see also][]{livio95}. Importantly they do not explicitly confirm the presence of non-spreading LTNR (``volcanoes'') that were speculated to exist by \citet{shara82}. 

Despite the lack of convincing observational evidence on the existence of LTNRs in AWDs, puzzling short-lived high-amplitude variations have been observed in a number of AWDs, most notably in the magnetic system TV Columbae \citep[TV Col:][]{schwarz88,hellier93b}. High ionization helium and nitrogen lines were observed to strengthen during these fast bursts and outflow velocities greater than $3500$ km s$^{-1}$ were observed during peak luminosity when P-Cygni profiles developed in UV spectral lines \citep[][]{szkody84}. At that time no clear explanation was found for such fast variability as well as the outflowing velocities.

The {\sl Transiting Exoplanet Survey Satellite } (\textit{TESS}) has drastically changed the observational status. Its unprecedented monitoring of the optical sky has yielded a number of AWDs where short-duration, fast-rise-exponential-decay events lasting a few hours grouped in pairs or triples, and with recurrence times of days to months have been observed. \citet{scaringi22} noted these events in the systems TV Col, EI UMa and ASASSN-19bh and \citet{SPZ22} noted similar bursts in the recurrent nova V2487 Oph during quiescence. \citet{scaringi22} conclude that these bursts are of thermonuclear origin, based on the energetics, the rise-time and the close resemblance to Type I X-ray bursts in accreting neutron stars, referring to them as \textit{micronovae}. \citet{SPZ22} instead come to the conclusion that the bursts are caused by magnetic reconnection events in the accretion disc possibly due to an extremely active companion.

Here we elaborate and expand in more detail on the possible thermonuclear origin of these fast bursts in AWDs which may give rise to micronovae events as proposed in \citet{scaringi22}. In Section\ \ref{sec:model} we introduce a model where the surface magnetic fields of AWDs may confine the flow of accreted material on to the WD surface and allow the pressure at the base of the accretion column to reach the critical pressures required to initiate localised TNRs. Section\ \ref{sec:discussion} discusses some possible limitations to the model and discusses our results in light of the observations of \citet{scaringi22} and \citet{SPZ22}.

\section{Magnetically confined accretion}
\label{sec:model}

The model we propose allows an accretion column on the magnetic poles of AWDs to be confined by the WD magnetic field and to grow in mass over time. As this happens the pressure exerted on to the WD due to the column's weight causes the column base to sink to larger depths. If this magnetic confinement can hold until the pressure at the base of the accreted column reaches $P_{\rm crit}\approx 10^{18}$ dyn cm$^{-2}$, a TNR may start \citep[e.g.][]{bode,jose}. The ignition burns through most of the overlaying accumulated mass in the column. The process can repeat every time the pressure at the column's base reaches the required pressure to drive a TNR.

\begin{figure*}
    \includegraphics[width=1.95\columnwidth]{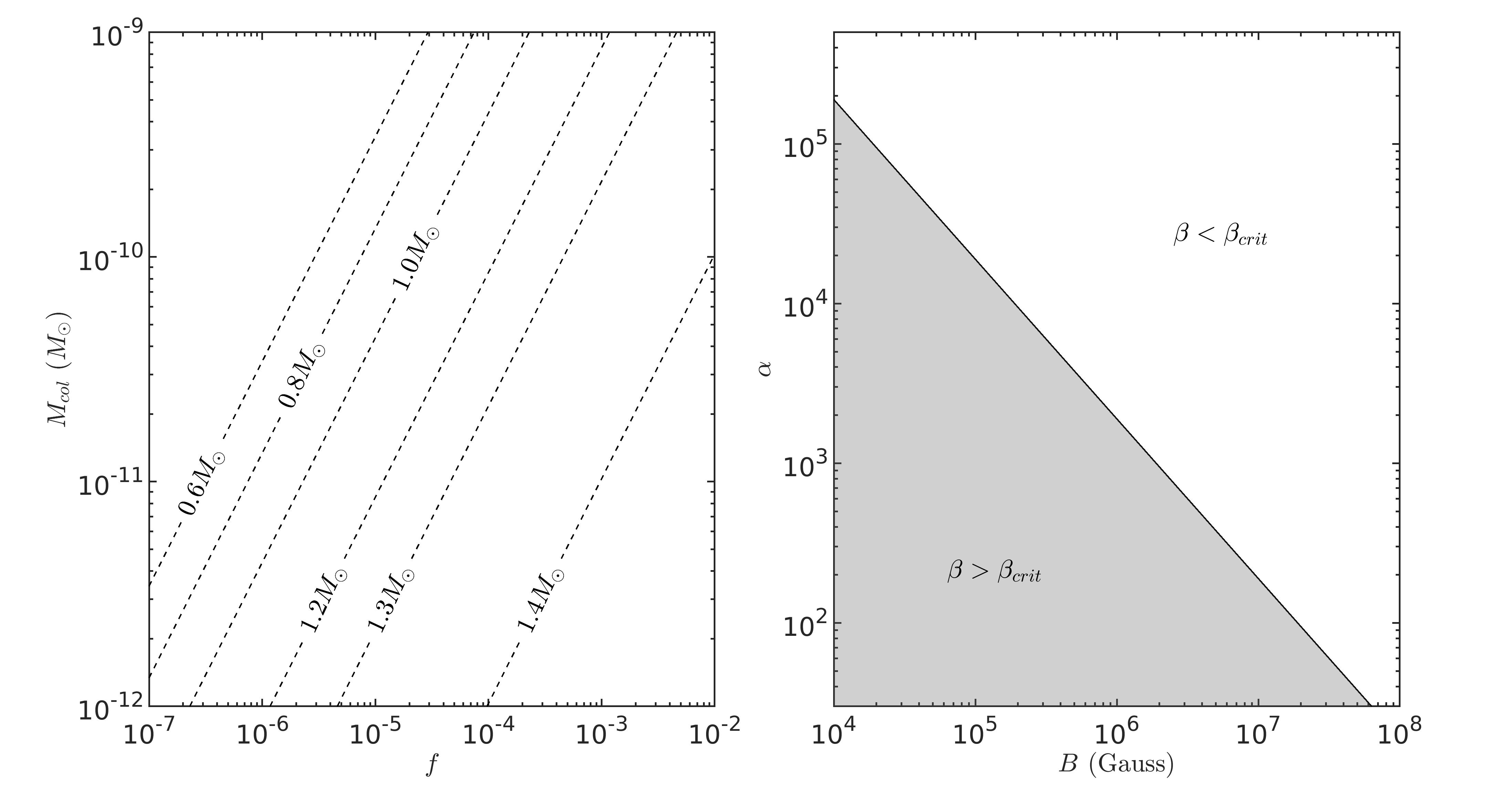}
    \caption{Left-panel: Range of column masses ($M_{col}$) required to reach $P_{base} \approx P_{crit} \approx 10^{18}$ dyn cm$^{-2}$. The plot has been computed with a range of WD masses as indicated by the dashed lines. Right-panel: Constraint on keeping the accretion column magnetically confined up to $P_{crit}=10^{18}$ dyn cm$^{-2}$. Gray shaded region shows regions where the column pressure will be too high and break the magnetic confinement.}
    \label{fig:1}
\end{figure*}

In magnetic AWDs with surface magnetic field strengths of $B \approx10^6$ G to $10^7$ G, the accretion flow impacts on the WD magnetic poles and remains confined by the magnetic pressure $P_{B} = \frac {B^2}{8\pi}$. This only happens if the ratio 
\begin{equation}
\beta = \frac{P_{\rm gas}}{P_{B}},    
\end{equation}
where $P_{\rm gas}$ is the gas pressure of the magnetically confined material, does not exceed a critical $\beta_{\rm crit}$. In general, for magnetically channelled accretion on to WDs, $P_{\rm gas}$ is substantially lower than $P_{\rm B}$. As the weight of the column grows over time the pressure at the base of the magnetically confined column ($P_{\rm base}$) also grows, and the column pressure being exerted radially downwards can translate azimuthally and begin exerting pressure on to the magnetically confined boundary. In this limit the requirement for material in the accretion column to remain confined becomes 
\begin{equation}\label{eq:magConf}
\beta = \frac{P_{\rm base}}{P_{B}}<\beta_{\rm crit}.
\end{equation}
If $\beta>\beta_{\rm crit}$ then the column pressure substantially distorts the magnetic field lines, and the accretion column may spread on to the WD surface. 

Given a net positive mass-accretion rate, into the magnetically confined accretion column, $\dot{M}_{\rm acc}$ (which takes into account any mass leakage from the column itself), the column mass grows with time $t$ such that 
\begin{equation}
M_{\rm col}(t)=\dot{M}_{\rm acc} t.
\end{equation}
Assuming an accretion column with a circular footprint area of radius $R_{\rm col}$, the fractional impact area on to the WD surface can be defined as \begin{equation}
f=\left( \frac{R_{\rm col}}{2R_{\rm WD}} \right) ^2,
\end{equation}
where $R_{\rm WD}$ is the WD radius. Using this definition, the pressure exerted at the base of the accretion column will grow such that 
\begin{equation}\label{eq:Pbase}
    P_{\rm base}(t) =  \frac{{GM_{\rm WD}M_{\rm col}(t)}}{4\pi f R_{\rm WD}^4},
\end{equation}
where $M_{\rm WD}$ is the WD mass and $G$ the gravitational constant. Over time the column mass will grow, and numerical magneto-hydrostatic results \citep[][]{HL85} have shown that the accretion column will remain confined by the magnetic pressure $P_B$ as long as 
\begin{equation}\label{eq:magConf2}
\beta(t) = \frac{P_{\rm base}(t)}{P_B} < \beta_{\rm crit}
\end{equation}
where 
\begin{equation}
    \beta_{\rm crit} \approx 7\alpha^2
\end{equation}
and
\begin{equation}
    \alpha=\frac {R_{\rm col}} {h},
\end{equation}
with $h$ being the height of the accumulated material in the column \citep[see][]{HL85}. If $M_{\rm col}(t)$ is able to become large enough such that the column pressure is equal to or larger than $P_{\rm crit}\approx 10^{18}$ dyn cm$^{-2}$ \citep[e.g.][]{jose,yaron} while remaining magnetically confined (Eq. \ref{eq:magConf2}), then it is reasonable that a TNR may start. When this happens, the TNR, while heating up, may expand along the path of least resistance which in this case is along the lower pressure magnetically confined column material above it, eventually escaping from the WD surface, and burning most of the column mass $M_{\rm col}(t)$ in the process. It may also happen that the radiation pressure or turbulent dynamics in the column generated by the hot TNR will break the magnetic confinement, in which case the column material could spread on to the WD surface as it is being ignited. Nonetheless, after the localised TNR takes place, the process of accumulating mass in a magnetically confined accretion column will restart, and another micronova will then be observed with a recurrence time of
\begin{equation}
    t_{\rm rec} = \frac{M_{\rm col}}{\dot{M}_{\rm acc}}.    
    \label{equ:trec}
\end{equation}
The left panel Fig. \ref{fig:1} shows the column mass required to reach $P_{\rm base}\approx P_{\rm crit} \approx 10^{18}$ dyn cm $^{-2}$ as a function of fractional accretion area for AWDs with masses in the range $0.6M_{\odot}$ up to $1.4M_{\odot}$. In making the figure we used a mass-radius relation to determine the WD radius \citep[][]{MRrelation}, and used Eq. \ref{eq:Pbase} to compute $P_{\rm base}$. The right panel of Fig. \ref{fig:1} shows the minimum $\alpha$ required to maintain the accretion column magnetically confined at least until $P_{\rm crit}=10^{18}$ dyn cm $^{-2}$ is reached. Both panels show that the magnetic confinement expected from magnetic AWDs ($B>10^6$ G) should be enough to build enough pressures at the bases of accretion columns to initiate localised TNRs. 

Assuming the material being burned during micronovae is freshly accreted hydrogen from the companion donor star, the CNO cycle flash will yield $\approx 10^{16}$ erg g$^{-1}$ \citep[e.g.][]{starrfield71,starrfield72,starrfield76}, and we can adopt this value to convert the radiated energy during micronovae into equivalent column masses $M_{\rm col}$. Micronovae have been observed to release between $10^{38}$ erg up to $10^{39}$ erg \citep[][]{scaringi22,SPZ22}, $\approx10^{6}$ times less than the energies released in classical novae (thus the term {\sl micronova} describing these events). This then translates to column masses in the range $5\times10^{-12}M_{\odot} < M_{\rm col} < 5\times10^{-11}M_{\odot}$. For AWDs with $M_{\rm WD}\approx0.8M_{\odot}$ \citep[typical for AWDs:][]{Z11,pala21} the corresponding fractional accretion area required by the model would then be $f\approx10^{-6}$. Although this value is low when compared to those inferred from X-ray observations of other magnetic AWDs \citep[][]{Hellier97,LM19}, it is still allowed by models of magnetically channeled accretion flows where material precipitates on to the polar cap in discrete filaments \citep[][]{king95,FKR}. However AWDs with masses of $M_{\rm WD}\approx1.3M_{\odot}$ are able to achieve the required pressures with accretion fractional areas of $f\approx10^{-4}$, increasing to $f>10^{-3}$ for WDs approaching the Chandrasekhar limit. These inferred accretion fractional areas are consistent with those observed with at least one of the systems displaying micronovae \citep[TV Col: ][]{LM19}. However, the large mass required to achieve $f\approx 10^{-4}$ in TV Col appears inconsistent with that inferred from X-ray observations of $0.74M_{\odot}$ \citep[][]{LM19}. It is interesting to further note that typical mass-transfer rates of $\dot{M}_{\rm acc}=10^{-10} M_{\odot}$yr$^{-1}$ for AWDs will achieve $P_{\rm base}=P_{\rm crit}$ with a recurrence time of $t_{\rm rec}\approx100$ days. The energy release and recurrence times match the observations of micronova in \cite{scaringi22}. On the other hand, a mass-accretion rate of  $\dot{M}_{\rm acc}=10^{-8} M_{\odot}$yr$^{-1}$ will yield $t_{\rm rec}\approx 1$ day, which also qualitatively matches the observations of the RN V2487 Oph \citep[][]{SPZ22}.

There is perhaps a further mechanism which may allow some of the accreted material to reach $P_{crit}$ with wider accretion fractional areas $f$ (and/or alternatively shorter timescales than Eq. \eqref{equ:trec}). If as the column mass grows over time the density at the base of the column becomes comparable to or higher than that of the underlying WD, then this configuration may lead to a Rayleigh–Taylor instability. If and when this happens, freshly accreted column material may be brought at deeper depths reaching $P_{crit}$ and triggering a micronova. This process may also lead to partial burning of the accreted fuel, which would then completely burn in sets of smaller bursts as observed in TV Col or EI Uma. Whether high enough column densities can be reached to surpass those of the underlying WD will depend crucially on column temperature profile. This in turn will depend on the conduction, dissipation from the column walls, and accretion rate, all of which seem to favour high accretion rates to maintain higher column temperatures. Whether the conditions for the instability are reached (i.e. a significant density gradient between the column base and the underlying WD and up to what depth the instability will develop) should be assessed by magneto-hydrodynamic numerical calculations.

\section{Discussion and Conclusion} \label{sec:discussion}

Although the model described in Section \ref{sec:model} shows that localised TNRs are possible if the flow of material can remain magnetically confined, it is important to comment on some of its limitations. First the model assumes that the WD magnetic field lines are solidly anchored at the bottom of the accretion column. This in turn provides the magnetic confinement required for the column to grow in mass over time. Because the exterior layers of WDs are not solid, it is possible that some lateral spreading of the column material on to the surface does occur. In particular, some of the various magnetised plasma instabilities might be at play. We specifically refer to $\dot{M}_{\rm acc}$ as the \textit{net} column mass-accretion rate, which does not necessarily have to be the same as the mass-accretion rate from the disc on to the WD. This is because some of the material may experience either lateral spreading or be accreted outside of the magnetically confined column (or both). In this case we would still expect micronovae to occur, but with longer recurrence times than those computed in Section \ref{sec:model}. Also important to address are the effects of the column settling into the WD, because this may decrease the column mass. If the settling time scale is faster than what can be accreted through $\dot{M}_{\rm acc}$, then a micronova may be inhibited because the column does not grow in mass over time. Given this consideration the model thus appears to favour systems with mass-accretion rates that are higher than the settling time scale of AWDs. Further modelling to include the effects of settling will us allow to determine whether this is the case. 

The depth where a magnetically confined TNR occurs may determine how much radiation and its associated wavelength from the TNR escapes to reach observers. Deeper TNRs may yield fainter and redder micronovae than those occurring closer to the WD surface. In this respect it is important to note that higher-mass WDs achieve $P_{\rm base}\approx P_{\rm crit}$ closer to their surfaces than lower-mass WDs. Thus higher-mass WDs not only provide more reasonable accretion fractional areas, but may also allow more radiation to escape because the TNR occurs at shallower depths. The model in Section~\ref{sec:model} also appears to disfavour magnetic AWDs with relatively low surface magnetic field strengths ($B<10^6$G) since these systems require columns that are extremely short and wide with $\alpha>10^4$ to be able to confine the column magnetically (see Fig. \ref{fig:1}, right panel).

One may ask why the micronovae do not ignite the whole WD outer layer leading to a classical nova explosion. If, following the trigger of a localised TNR, the hot fluid is ejected following the magnetically confined boundary, the heat may be dissipated outside of the WD. Depending on the surface composition of the WD, the temperature reached by the outer layers may also be too low to trigger unstable burning. Observations of TV Col, for example, during one of its micronova events show clear evidence of fast outflows only during the peak of the bursts \citep[][]{szkody84}, suggesting the ejection of material is driven by the micronovae themselves. If this material is part of the burning column material, this would act as a substantial sink of heat.

Finally, it is important to comment on the reasons why some magnetic AWDs appear to display micronovae while some do not, and why at least one system displays both classical novae and micronovae. Following the simple model in Section \ref{sec:model}, the requirement to trigger a micronova is that $P_{\rm base}\approx P_{\rm crit}$, and this is achieved with small accretion fractional areas for lower-mass WDs in order to explain the observed energies released. It is thus possible that AWDs displaying micronovae have large WD masses. This would provide reasonable accretion surface areas consistent with observations. However the WD masses required to obtain reasonable accretion surface areas appear higher than those inferred from X-ray observations \citep[][]{LM19}. In some systems it may further be that the combination of mass-transfer rate, surface magnetic field, WD spin, and spin-to-orbit alignment provides unfavorable conditions to achieve $P_{\rm base}\approx P_{\rm crit}$ within $t_{\rm rec}$ as defined in Eq. \ref{equ:trec}. This situation can also be further complicated if the impact area of the magnetically confined accretion flow varies over time. For example, the high mass-transfer rate system V2487 Oph could accrete outside of its magnetically confined region for most of the time, either because of changes in mass-accretion rate or other factors relating to where material latches on to the magnetic field lines. In this case fresh material would spread on to the WD and accumulate mass in preparation for the next global nova eruption. If and when material is able to remain magnetically confined to a small enough fractional area for at least $t_{\rm rec}$ then we may observe micronovae.  

Among the four AWDs confirmed so far to display micronovae, two (EI UMa and TV Col) belong to the magnetic class of intermediate polars \citep[IPs:][]{thorstensen86,hellier93a}. V2487 Oph has been suggested to harbour a magnetic AWD although no coherent pulsations have been detected so far \citep[][]{heranz02}, while ASASSN-19bh is a recently identified system with a suspected magnetic WD accretor \citep[][]{scaringi22}. The orbital periods of EI UMa (6.4\,h) and TV Col (5.5\,h) are long and V2487 Oph is a recurrent nova with a likely orbital period of 1.2\,d \citep[][]{SPZ22}, which all point to high mass-accretion rate systems. Therefore magnetically confined TNRs in these AWDs appear to be a feasible mechanism. Furthermore, the RN V2487 Oph is expected to harbour a high-mass WD and a high-mass accretor has also been considered to explain the detected large positive superhumps in TV Col \citep[][]{retter03}, and by extension those detected in EI Uma \citep[][]{scaringi22}. We have also found reports in the literature of two further systems that appear to have displayed a micronova in the past. The IP V1223 Sgr (orbital period of 6.4\,h) has been observed to display a single burst lasting several hours \citep[][]{amerongen89}, while three bursts each lasting less than a day with a recurrence of about $60$d have been observed in the IP DW Cnc (orbital period of 86.1\,m) during a high accretion state \citep[][]{duffy22}. The model presented in Section \ref{sec:model} requires relatively small accretion fractional areas on to the AWD, as well as relatively high mass-accretion rates and WD masses. Future observations of the systems mentioned here, especially in between successive micronova events (and specifically at X-ray wavelengths) will allow us to further test these expectations.

While different mechanisms explaining the rapid bursts have been proposed \citep[][]{SPZ22,shara82}, a quantitative model has not yet been developed. The model presented here has the potential to explain both the observed burst energies and recurrence timescales. Detailed time dependent magneto-hydrodynamic simulations of accretion columns in magnetic AWDs are required to further test the model presented here as a mechanism to trigger micronovae. Further multi-wavelength and long-term monitoring of micronovae will also be crucial in testing the model and in identifying the true origin of micronovae.

\section*{Acknowledgements}
The authors wish to thank F.X. Timmes and D. Koester for useful exchanges during the preparation of this manuscript, as well as Christopher Tout for the usefull and timely review of the manuscript. P.J.G. is supported by NRF SARChI grant 111692. D.dM. acknowledges financial support from the Italian Space Agency (ASI) and National Institute for Astrophysics (INAF) under agreements ASI-INAF I/037/12/0 and ASI-INAF n.2017-14-H.0 and from INAF ``Sostegno alla ricerca scientifica main streams dell’INAF'', Presidential Decree 43/2018 and from INAF ``SKA/CTA projects'', Presidential Decree 70/2016 and from PHAROS COST Action N. 16214. J.-P.L. was supported in part by a grant from the French Space Agency CNES. M.E.C. acknowledges NASA grants 80NSSC17K0008 and 80NSSC20K0193 and the University of Colorado Boulder. 

\section*{Data Availability}
There are no new data associated with this article.



\bibliographystyle{mnras}
\bibliography{magConfMicronova} 





\bsp	
\label{lastpage}
\end{document}